\documentclass[%
 reprint,
 amsmath,amssymb,
 aps,
 prab,
floatfix,
]{revtex4-2}

\usepackage{graphicx}
\usepackage{dcolumn}
\usepackage{bm}
\usepackage{siunitx}
\usepackage[utf8]{inputenc}

\pdfoutput=1

\begin{document}

\preprint{APS/123-QED}

\title{Precision alignment and tolerance of a plasma wakefield accelerator in a laser-ionized plasma source}

\author{Valentina Lee$^1$}
\author{Robert Ariniello$^2$}
\author{Douglas Storey$^2$}
\author{S{\'e}bastien Corde $^{2,3}$}
\author{Claudio Emma$^2$}
\author{Spencer Gessner$^2$}
\author{Mark Hogan$^2$}
\author{Alexander Knetsch$^2$}
\author{Nathan Majernik$^2$}
\author{Brendan O'Shea$^2$}
\author{Ivan Rajkovic$^2$}
\author{Michael Litos$^1$}

\affiliation{$^1$Center for Integrated Plasma Studies, Department of Physics, University of Colorado Boulder, Boulder, Colorado 80309, USA}
\affiliation{$^2$SLAC National Accelerator Laboratory, Menlo Park, California 94025, USA}
\affiliation{$^3$LOA, ENSTA, CNRS, Ecole Polytechnique, Institut Polytechnique de Paris, F-91762 Palaiseau, France}

\date{\today}

\begin{abstract}

We present a novel method for aligning a laser ionized plasma source to a pair of ultra-relativistic electron beams that comprise a plasma wakefield accelerator (PWFA). We achieve alignment by analyzing the plasma afterglow light observed at two longitudinal locations as the plasma column is scanned across the electron beam. By analyzing the relative plasma light intensity at the two locations, we aligned an 85-cm plasma source to a \qty{10}{GeV}, \qty{1.6}{nC} electron beam to within \qty{10}{\micro\meter} offset, \qty{10}{\micro rad} tilt. The alignment is verified by analyzing the drive beam energy loss, energy transfer efficiency, and the witness beam energy gain as a function of the misalignment between the beams and the plasma. From this measurement, we extract the alignment tolerance required between the laser-ionized plasma source and electron beams, an important metric necessary for collider design studies or light source applications based on a laser-ionized plasma source.  
\end{abstract}

\maketitle


\section{\label{sec:intro}Introduction}

Plasma wakefield accelerators (PWFA) have shown orders of magnitude higher accelerating gradients compared to conventional radiofrequency (RF) accelerators, making them a lower-cost, more energy-efficient solution for future colliders~\cite{Litos2014, Litos2016, Blumenfeld2007, roser2023feasibility}. In an electron beam-driven PWFA, a “drive” bunch of electrons expels plasma electrons from the beam axis as it travels through a plasma, creating a plasma wake behind it. The much heavier ions remain stationary and attract the expelled electrons back to the axis, forming an ion bubble surrounded by a sheath of electron current. This structure simultaneously provides a transverse focusing force and a longitudinal accelerating field for a second “witness” electron bunch.

PWFA has been demonstrated using beam-ionized alkali vapor ovens~\cite{muggli1999photo, hogan2005multi}, beam-ionized noble gases~\cite{storey2024}, discharge plasma sources~\cite{lindstrom2021energy, lindstrom2024emittance, pena2024energy}, and laser ionized plasma sources. Beam-ionized alkali plasma sources based on heat pipe ovens offer the advantage of being self-aligning to the electron beam but suffers from head erosion of the drive beam~\cite{an2013} and overheating from excess deposition of beam energy, making them unsuitable for high repetition rate. Pre-ionized discharge plasma sources avoid head erosion and have been used to demonstrate energy spread~\cite{lindstrom2021energy} and emittance~\cite{lindstrom2024emittance} preservation with relatively low energy electron beams. However, both a discharge plasma source and an alkali vapor oven present challenges shaping the plasma density profile and offer limited optical access for optical diagnostics~\cite{lee2024} and laser injection schemes~\cite{ullmann2021all}.

Laser-ionized plasma sources can create plasmas with controlled longitudinal density profiles, required for emittance preservation~\cite{Floettmann2014, Xu2016, Ariniello2022}, and controlled diameter, necessary for advanced PWFA applications such as positron acceleration~\cite{Diederichs2019} and the ion channel laser~\cite{litos2018ICL}. A major challenge of this scheme is to align the plasma column to the electron beam with sufficient precision. This challenge is exacerbated with high-current electron beams, as it is difficult to accurately determine the electron beam vector at the waist using optical transition radiation-based (OTR) beam diagnostics due to the presence of camera-blinding coherent transition radiation (COTR). Traditional OTR-based alignment can usually achieve a plasma-to-beam (laser-to-beam) alignment of \qty{200}{um}.

For our application, at the Facility for Advanced Accelerator Experimental Tests II (FACET-II) at the SLAC National Accelerator Laboratory~\cite{Yakimenko2019}, a \qty{85}{cm} long, \qty{200}{\micro\meter} diameter laser-ionized plasma is formed in a \qty{1.35}{torr} hydrogen gas (plasma density of \qty{4.5e16}{cm^{-3}}). The diameter of the wake is approximately $\lambda_p=\qty{150}{\micro\meter}$, necessitating a plasma-to-beam (laser-to-beam) alignment of better than \qty{50}{um} along the length of the plasma. 

In this paper, we describe the experimental setup in Sec.~\ref{Exp} and introduce a novel alignment technique based on plasma afterglow light, that is suitable for the high current FACET-II beams, in Sec.~\ref{Alignment}. This technique enables us to align the plasma to the electron beam with a precision of \qty{10}{\micro rad} in tilt and \qty{10}{\micro\meter} in offset. Following the alignment, we explore the plasma-to-beam alignment tolerance for this experimental setup in Sec.~\ref{tolerance}.

\begin{figure*}
\centering
\includegraphics[width=2\columnwidth]{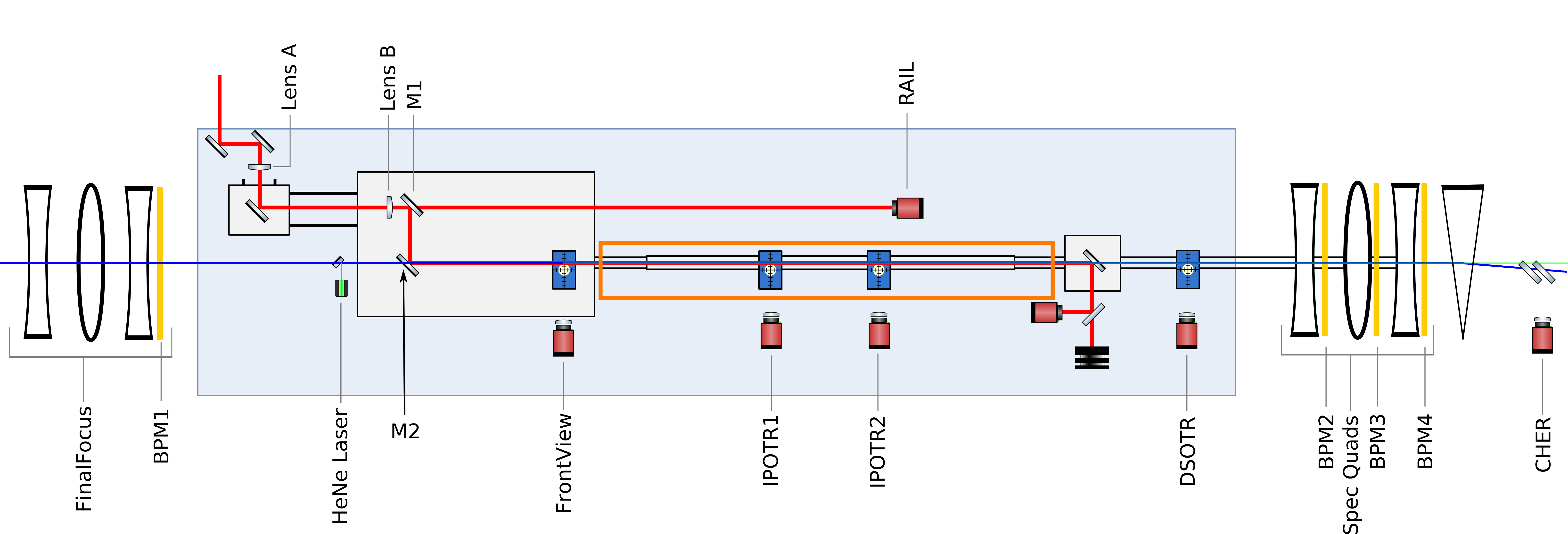}
\caption{\label{fig:IPArea} Experimental setup around the interaction point (IP area) at FACET-II. The electron beams travel from left to right (blue line) and are focused by the Final Focus, which is composed of three quadrupoles. Along the electron beam path, there are four insertable OTR screens, monitored by cameras: FrontView, IPOTR1, IPOTR2, and DSOTR, respectively. The distance between IPOTR1 and IPOTR2 is \qty{40}{\cm}. The electron beam waist is located within the orange area. The compressed laser beam (red line) enters the IP area from the top of the figure and passes through the tandem lens pair (lenses A and B). The laser beam is then reflected off mirrors M1 and M2, which are used to align the laser beam to the electron beam vector. The laser intensity incident on M2 has a donut-shaped profile, and because M2 is a hole mirror, the electron beam passes through its center. The laser focus appears within the orange area. Mirrors can be inserted at a 45$^\circ$ angle at IPOTR1 and IPOTR2 to view the laser focus. An alignment laser (green line) is used to mark the electron beam vector in the IP area. Finally, the electron beams are imaged by the Spec Quads to a Cherenkov based detector (CHER). Four beam position monitors (BPMs) are marked behind quadrupoles.}
\end{figure*}

\section{\label{Exp}Experiment}

At FACET-II, the plasma is formed by a \qty{200}{mJ}, \qty{50}{fs} laser pulse that passes through a pair of diffractive optics, denoted Lens A and Lens B in Fig.~\ref{fig:IPArea}, forming an \qty{85}{\cm} long Bessel focus with a focal spot diameter of \qty{120}{\micro\meter} beginning \qty{1.45}{m} downstream of Lens B. When the beamline is filled with a \qty{4.5e16}{cm^{-3}} hydrogen gas, the laser is modestly refracted by the partially ionized plasma formed at the head of the pulse resulting in a \qty{200}{\micro\meter}-wide plasma column. The plasma column vector follows the ionization laser's trajectory through the focusing lenses. In between Lens B and the start of the Bessel focus, the laser reflects off two mirrors (M1 and M2) that are used to adjust the alignment of the laser relative to the electron beam. In addition to mirrors M1 and M2, both Lens A and Lens B are located on 2-axis translation stages allowing the plasma to be translated in the transverse plane independent of the trajectory angle. The leakage light from M1 is viewed by the RAIL camera to collect shot-to-shot laser position at focus. 

The FACET-II linac delivers \qty{10}{GeV} electron beams, in a single bunch configuration, at \qty{10}{Hz}, with a total charge of \qty{1.6}{nC}, transverse root-mean-square (RMS) spot sizes of \qty{20}{\micro\meter} $\times$ \qty{20}{\micro\meter}, bunch length of \qty{20}{\micro\meter}, resulting in a peak current of $>$\qty{30}{kA}, and a normalized emittance in both transverse dimensions of approximately \qty{10}{mm-mrad}~\cite{yakimenko2019facet}. A two-bunch configuration is achieved by inserting a collimator into the center of the beam in a dispersive region~\cite{clarke2012facet}, creating a drive beam of ~\qty{0.9}{nC} and a witness beam of ~\qty{0.1}{nC}. 

The electron beam travels from left to right in Fig. \ref{fig:IPArea}, focused by a set of three quadrupoles to produce an waist with adjustable beta function of \qty{10}-\qty{50}{cm} with an adjustable waist location around the entrance of the laser-ionized plasma column at the interaction point (IP). The electron beam is timed to arrive \qty{2}{ps} behind the laser pulse. At the IP, four viewing ports are equipped with cameras (FrontView, IPOTR1, IPOTR2, and DSOTR) to provide views of the beam axis from the side. The middle two cameras (IPOTR1 and IPOTR2) view the region with laser ionized plasma. Metal foils can be inserted into the beam line at \qty{45}{\degree} to the beam axis to serve as OTR-based beam position diagnostics at FrontView, DSOTR, IPOTR1, and IPOTR2. At the middle two ports (IPOTR1 and IPOTR2), mirrors can also be inserted at \qty{45}{\degree} to reflect the laser beam to the camera in order to diagnose its position at these locations.

Upstream of the IP, a low-power HeNe laser assists in marking the electron beam vector. Downstream of the IP, an electron imaging spectrometer, consisting of three quadrupoles and a dipole, transports and disperses the beam, to image the plasma exit plane onto a Cherenkov-based detector screen (CHER).  Beam position monitors (BPMs) are located before and after the IP to measure the shot-to-shot beam position before and after the beam-plasma interaction. Further details of the experimental area and the available diagnostics at FACET-II are provided in Ref.~\cite{storey2024}. 

\section{\label{Alignment}Plasma Afterglow-based Alignment Technique}

\begin{figure}[h]
\centering
\includegraphics[width=0.95\columnwidth]{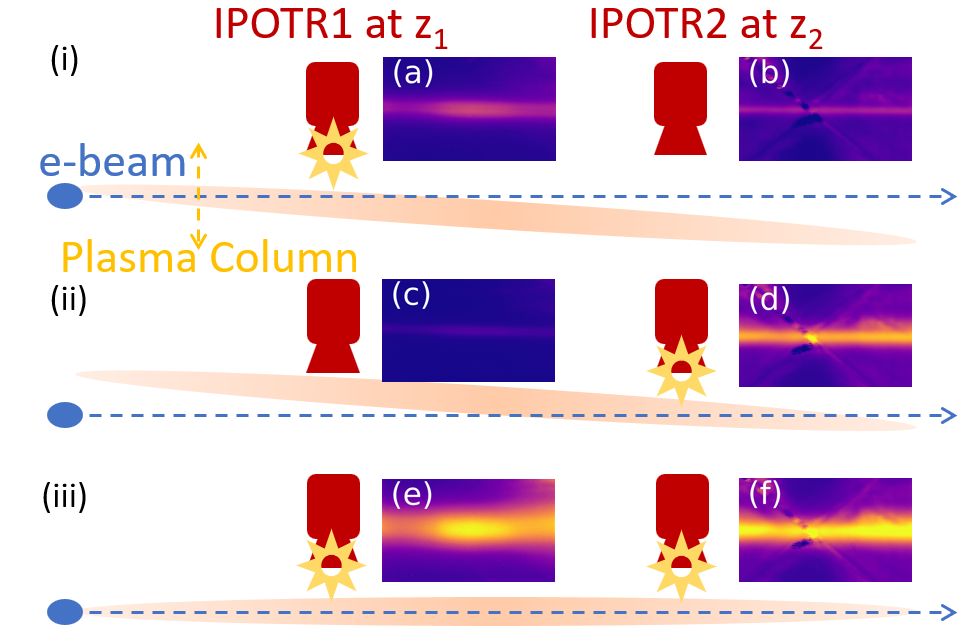}
\caption{\label{fig:Cartoon} Alignment method. The electron beam vector is shown as a blue dashed line, moving from left to right. The plasma column is scanned transversely across the beam vector. If the plasma column is tilted relative to the electron beam vector, IPOTR1 (a) initially observes more light than IPOTR2 (b). As the plasma moves past the electron beam, the light on IPOTR1 is reduced (c) while the light on IPOTR2 increases (d). When the plasma is aligned to the electron beam, both cameras observe bright plasma at the same time (e and f).}
\end{figure}

In this section, we introduce a novel alignment technique to align a \qty{85}{cm} plasma to the electron beam based on the plasma afterglow light emitted by the plasma, suitable for achieving the alignment requirement demonstrated in Sec. \ref{tolerance}. In the absence of the electron beam, the afterglow light emitted from a thin, laser-ionized plasma source has been observed to scale with plasma density and electron temperature~\cite{lee2024}. When an electron beam drives a wake in the plasma, the brightness of the plasma light has been shown to scale linearly with the amount of energy deposited into the plasma~\cite{oz2004, boulton2022}. Further, the amount of energy deposited---and thus light emission---is reduced, but not eliminated, when the beam narrowly misses the plasma \cite{adli2016}. This beam-driven light emission is an order of magnitude larger than the light from the laser-ionized plasma alone. Because of this feature, plasma afterglow light has been used for alignment and synchronization~\cite{scherkl2022plasma, deng2019generation}. Figure \ref{fig:Cartoon} (d) and (c) provide examples of plasma light images, with and without an electron-driven wake.

\begin{figure}[t]
\centering
\includegraphics[width=0.95\columnwidth]{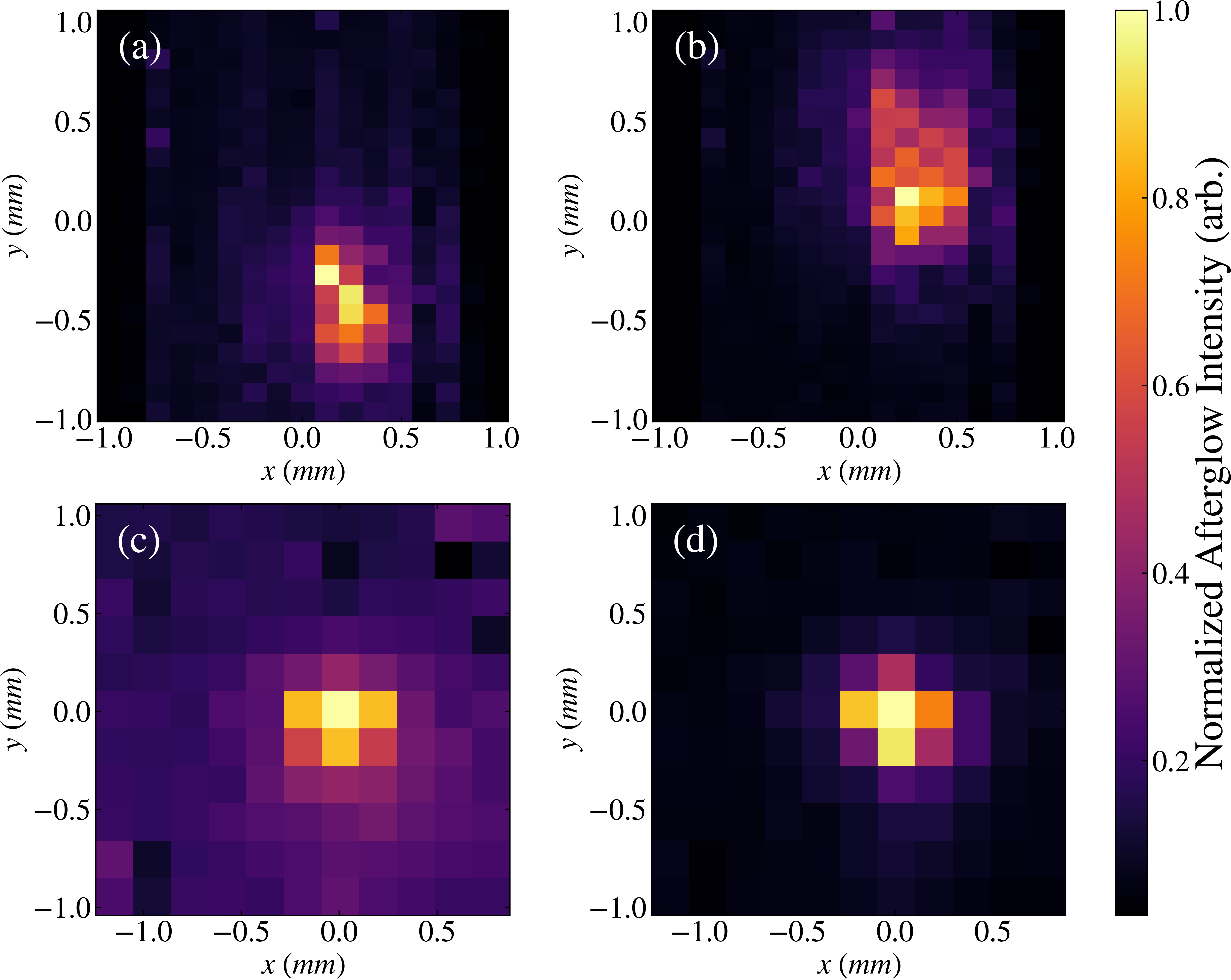}
\caption{\label{fig:RasterScan} Raster scan results. The color map indicates the normalized plasma afterglow intensity. (a) shows the plasma afterglow intensity as a function of the plasma column position relative to a defined beam axis. at IPOTR1 and (b) shows it at IPOTR2. When the plasma column is misaligned with the electron beam vector, the plasma afterglow intensity is maximized at different relative positions for each camera. In contrast, in an aligned case, the plasma afterglow intensity is maximized simultaneously at IPOTR1 (c) and IPOTR2 (d).}
\end{figure}

The plasma afterglow light was recorded on IPOTR1 and IPOTR2 while scanning the transverse position of the plasma column across the electron beam vector, by moving Lens A and B transversely, as illustrated in Fig. \ref{fig:Cartoon}. When the plasma column is misaligned with respect to the electron beam vector, the plasma afterglow brightness reaches its maximum at different plasma column positions on IPOTR1 and IPOTR2 [Fig.~\ref{fig:RasterScan}(a, b), (c, d)]. We can then calculate the misalignment angle and align the plasma column accordingly by adjusting M2 and translating Lens A and B until the plasma afterglow light is maximized simultaneously at both cameras [Fig.~\ref{fig:RasterScan}(e, f)]. This method offers a significant advantage over conventional OTR-based alignment, as it directly measures the vector misalignment through the interaction between the electron beam and the plasma, providing a more accurate and reliable signal.



\begin{figure}[h]
\centering
\includegraphics[width=0.95\columnwidth]{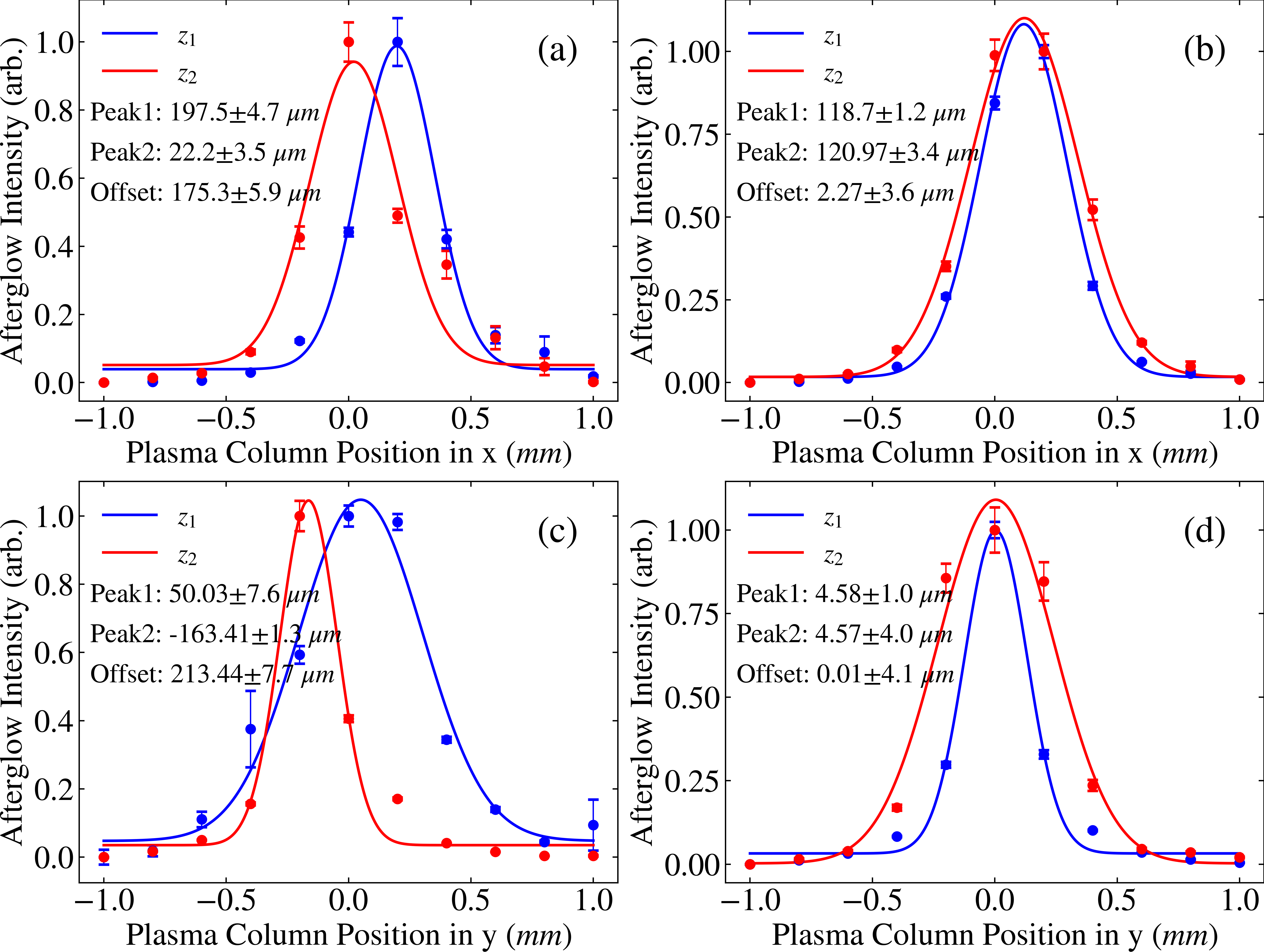}
\caption{\label{fig:1dScan} 1D scan results for plasma column alignment in both $x$ (a)(b) and $y$ (c)(d) directions. The plots show the normalized plasma afterglow intensity as a function of the plasma column position. (a) and (c) represent the results for OTR-based alignment (coarse alignment) at two different longitudinal positions: IPOTR1 ($z_1$) (blue) and IPOTR2 ($z_2$) (red); the peaks are separated by 175.3~$\mu m$ in the $x$ direction and 213.4~$\mu m$ in the $y$ direction. (b) and (d) correspond to the plasma afterglow-based alignment (fine alignment) results. The markers represent the experimental data points with error bars, and the curves are the Gaussian fitted results. As the alignment improves from coarse to fine, the peaks of the afterglow intensity curves at IPOTR1 and IPOTR2 converge to 2.27~$\mu m$ in $x$ and 0.01~$\mu m$ in $y$, indicating successful alignment of the plasma column with the electron beam.}
\end{figure}

The results of a two-dimensional scan of the laser (plasma) position are shown in Fig.~\ref{fig:RasterScan}. Although in theory this alignment procedure only needs to be performed once after the misalignment is measured, in practice an iterative process with a gain of 0.5 is used for two reasons. First, it better mitigates measurement noise. For example, if the electron beam possesses current spikes~\cite{storey2024} and ionizes the plasma, the resulting afterglow measurement may not accurately reflect the alignment; the iterative process ensures the alignment is achieved once the scan is converted. Second, the tip-tilt of M2 is controlled by piezo linear actuators (picomotors), which have repeatable linearity only for small steps. The calibration between angle and picomotor step becomes less accurate for large moves; therefore, a \qty{200}{\micro\meter} misalignment cannot be corrected accurately in a single step.

Acquiring the full 20×20-step transverse scan, with 10 shots per step as shown in Fig.~\ref{fig:RasterScan}, takes over an hour due to the network-related bottleneck when saving data. In practice, the process needs to be repeated several times to achieve alignment. As long as the initial alignment is within \qty{0.5}{mm}, iterative one-dimensional scans can be used in the vertical and horizontal axes to achieve the best possible degree of alignment in significantly less time than a high-resolution 2-dimensional scan. 

At FACET-II, we have implemented an automated procedure to perform this alignment. First, with the laser blocked, we insert metal foils at FrontView and DSOTR to record the electron beam trajectory using points far from the waist (to minimize the impacts of COTR). Next, we use the same two screens to align the HeNe alignment laser to the electron beam vector (see Fig.~\ref{fig:IPArea}). To obtain a coarse initial alignment of the ionization laser with the beam vector, we insert mirrors at IPOTR1 and IPOTR2 to view the laser beam position and use mirrors M1 and M2 for a two-point alignment. Next, we conduct two 1D translation scans of the ionization laser across the $x$ and $y$ axes while the e-beam is present. The angle of the laser is fixed during these scans. Typical measurements of plasma interaction strength after OTR-based alignment are shown in Fig. \ref{fig:1dScan} (a) and (b), showing a misalignment as large as \qty{200}{\micro\meter}, \qty{80}{\micro rad}.

Finally, we proceed with an automated 1D plasma afterglow-based alignment, iteratively measuring the angular misalignment by the afterglow signal in $x$ and $y$ and adjusting M1 and M2 to correct the misalignment until alignment is achieved in both directions. The results after the full alignment procedure are shown in Fig. \ref{fig:1dScan} (c) and (d). The residual peak separation from the Gaussian fits demonstrates alignment better than \qty{10}{\micro rad} of tilt and \qty{10}{\micro\meter} of offset. 

\section{\label{tolerance}Alignment Tolerance Study}
The geometric tolerance of a PWFA in a laser-ionized plasma source can be estimated as:
\begin{equation}
\label{eq:tor}
    dr+ L sin(d\theta)\leq 2 (R_p-r_b)
\end{equation}
where $R_p$ is the radius of the plasma column, $r_b$ is the blowout radius, $L$ is the length of the plasma, $dr$ is the transverse offset from the center at the start of the plasma column, and $d\theta$ is the angular misalignment. With $2R_p = \qty{200}{\micro\meter}$, $2r_b \approx \lambda_p = \qty{157}{\micro\meter}$ ($\lambda_p$ is the plasma wavelength), and $L = \qty{85}{cm}$, we can predict that $dr \leq \qty{43}{\micro\meter}$ when $d\theta = 0$, and $d\theta \leq \qty{50.6}{\micro rad}$ when $dr = 0$. This calculation seems straightforward. Previous studies have shown that a misaligned drive beam can be attracted into the plasma column~\cite{adli2016, muggli2001refraction}, which suggests that the alignment tolerance may be greater than that predicted by the simple geometric argument of Eq.\ref{eq:tor}, implying that slight misalignments can be “corrected” by the attractive force, thereby increasing the effective tolerance.

\begin{figure}[t]
\centering
\includegraphics[width=1\columnwidth]{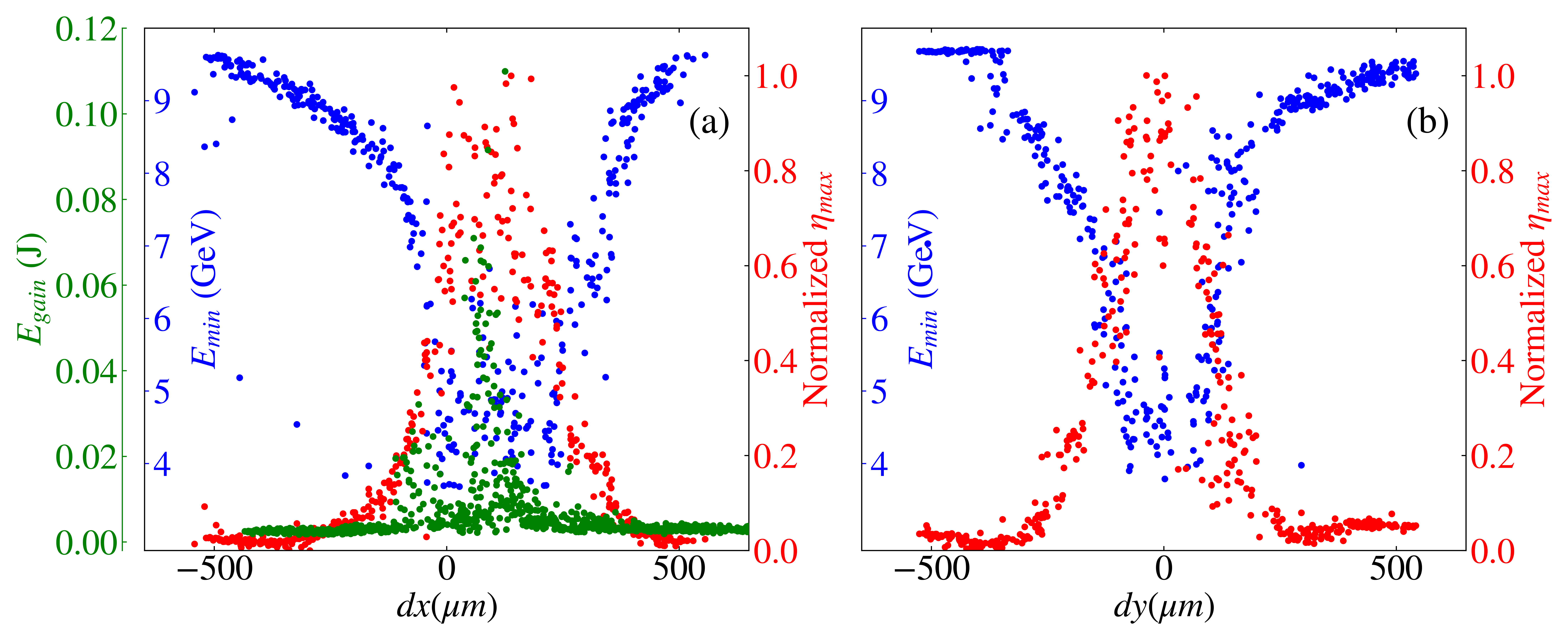}
\caption{\label{fig:PWFAPerf1} PWFA performance as a function of beam-to-plasma misalignment in $x$ (a) and $y$ (b). The markers represent the experimental data points of the minimum energy of the decelerated electrons ($E_{min}$, blue), the upper bound of beam-to-wake energy transfer efficiency ($\eta_{max}$, red), and the energy gained after the PWFA ($E_{gain}$, green). The $E_{gain}$ result shows that the transverse misalignment must be less than \qty{40}{\micro\meter}.}
\end{figure}

To study the alignment tolerance of the PWFA, we scanned the plasma transversely across the e-beam vector, similarly to the alignment method introduced in Sec.\ref{Alignment}. The shot-to-shot plasma-to-beam misalignment is measured using the laser far-field camera and the RAIL camera (to determine the shot-to-shot laser vector) and the two BPMs downstream of the IP (to determine the shot-to-shot beam vector)~\cite{lindstrom2020matching}. The energy spectra are analyzed, and the minimum energy of the decelerated electrons ($E_{min}$), the beam-to-wake energy transfer efficiency ($\eta_{max}$, is extracted from the spectra assuming the missing charge has the minimum energy visible on the screen), and the witness beam energy gain ($E_{gain}$) as functions of misalignment in both transverse directions are plotted in Fig.\ref{fig:PWFAPerf1}. 

The $E_{min}$ and $\eta_{max}$ distributions show a tolerance of $\approx$ \qty{150}{\micro\meter}, while a tolerance of $\approx$ \qty{40}{\micro\meter} is necessary to maintain a consistent value of $E_{gain}$. Although previous studies suggested that a misaligned drive beam may be pulled into the plasma column~\cite{adli2016}, the driven wake is noticeably weaker in this case, as the sheath electron current is smaller at the misaligned edge. Moreover, there is likely an offset between drive and witness, which reduces the size of the transverse acceptance window, relative to the drive. This effect is evident from the more limited tolerance observed in $E_{gain}$. This finding indicates that $E_{min}$ may overestimate the alignment tolerance of a PWFA in a laser-ionized plasma source. Instead, acceleration metrics such as $E_{gain}$ provides a more reliable indication. The results show that the transverse offset for the parallel plasma column must be less than \qty{40}{\micro\meter} = $0.93 \times 2(R_p-r_b)$ to maintain an $E_{gain}$ consistent within 97\% of the peak achievable value. 

\begin{figure}[t]
\centering
\includegraphics[width=0.9\columnwidth]{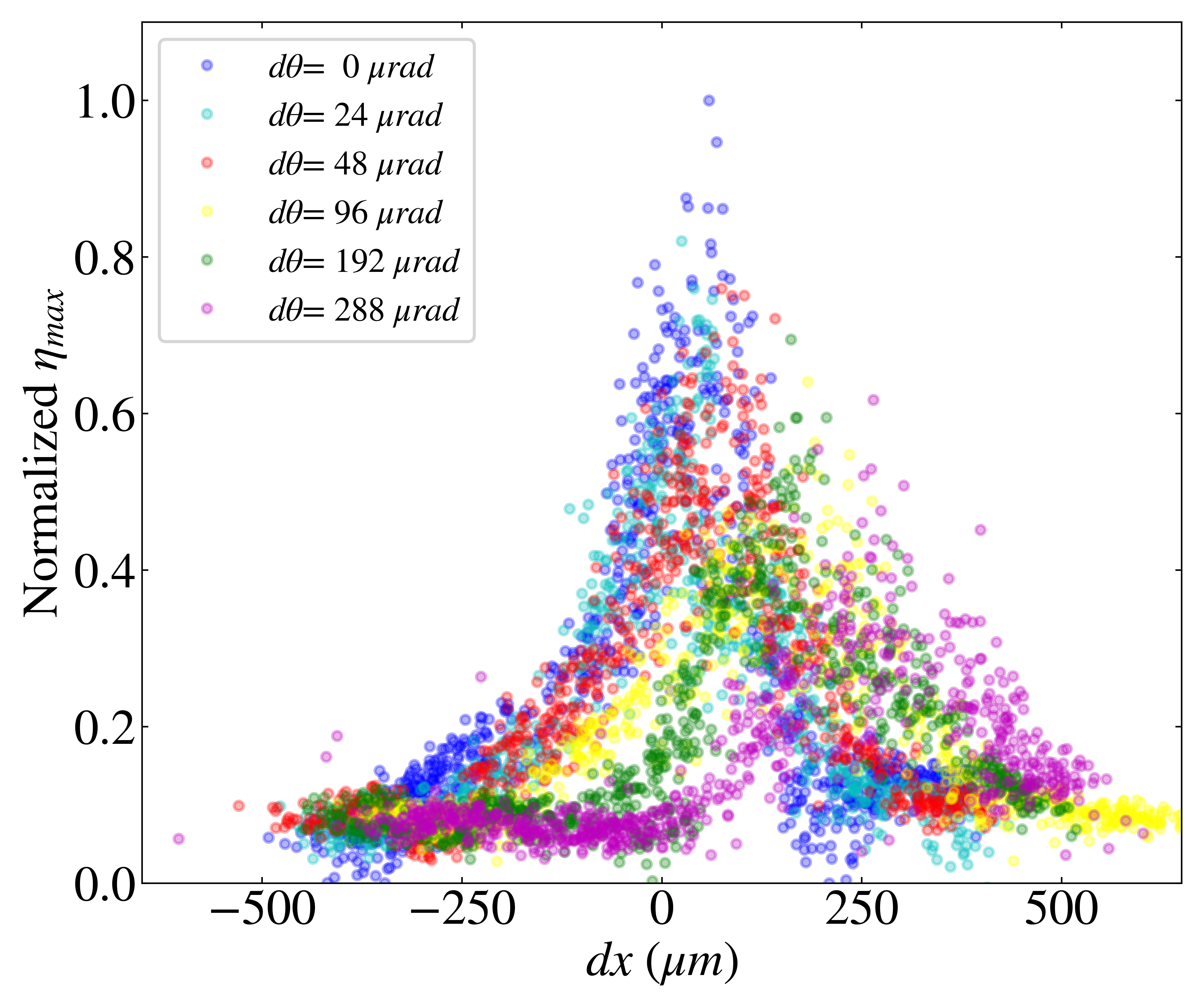}
\caption{\label{fig:PWFAPerf2} PWFA performance as a function of beam-to-plasma misalignment with different angular misalignment. The markers represent the experimental data points of the upper bound of beam-to-wake energy transfer efficiency ($\eta_{max}$). The results indicate that the angular misalignment tolerance lies between \qty{24}{\micro rad} and \qty{48}{\micro rad}: at \qty{24}{\micro rad} (light blue), the outcome is consistent with the aligned case (dark blue), whereas at \qty{48}{\micro rad} (red), the amplitude of the peak begins to decrease.}
\end{figure}

Figure~\ref{fig:PWFAPerf2} shows the beam-to-wake energy transfer efficiency ($\eta_{max}$) as a function of transverse and angular misalignment. At each step of the angular misalignment scan, the transverse position was scanned. The results show that the angular misalignment tolerance to maintain a consistent $\eta_{max}$ within 85\% of the peak achievable value is less than \qty{48}{\micro rad}. When normalized to the plasma length, $L$, this corresponds to an angular offset tolerance of \qty{48}{\micro rad} = 0.95 $\times$ 2($R_p-r_b$)/L.

Both the transverse and angular tolerances are consistent with the geometric calculation, indicating that geometric estimates are a reliable metric for determining alignment tolerance in a laser-ionized PWFA. At FACET-II, with a laser RMS pointing jitter of \qty{10}{\micro rad} and an electron beam position jitter of up to \qty{30}{\micro\meter} at its waist, stable PWFA operation is successfully demonstrated in a laser-ionized plasma source.

\section{\label{conclusions}Conclusions}
We have presented a novel alignment technique developed at FACET-II to precisely align a meter-scale laser-ionized plasma source with a 10 GeV, 1.6 nC high-current electron beam. This method automates two-point alignment based on plasma afterglow diagnostics. By performing 1D scans along the $x$ and $y$ axes, we measured and corrected misalignments, achieving alignment precision within \qty{10}{\micro\meter} offset and \qty{10}{\micro rad} tilt. The entire alignment could be achieved in 1 hour while operating at 10 Hz, limited primarily by data transfer and storage rates.

One limitation of this method is the dependency of plasma afterglow brightness on the wake strength driven by the beam, which can be influenced by beam ionization rather than by alignment accuracy to the laser-ionized plasma column. This issue can be mitigated by reducing the beam current through adjustments to beam compression, ensuring reliable alignment diagnostics, or by utilizing a high ionization threshold gas, such as helium.

Following successful plasma-to-beam alignment, we measured the alignment tolerance of the PWFA interaction by analyzing the minimum energy of the decelerated drive-beam particles ($E_{min}$), the drive beam-to-wake energy transfer efficiency $\eta_{max}$, and the energy gain of the witness beam ($E_{gain}$). Measurements of the drive beam performance via $E_{min}$ and $\eta_{max}$ showed a more relaxed alignment tolerance than that of the witness beam acceleration, $E_{gain}$, due to the attractive force exerted by the plasma column. Meanwhile, the geometric model predicted the tolerance of the witness beam acceleration via $E_{gain}$ very well. At FACET-II, the required level of precision has been achieved, enabling stable PWFA operation in a laser-ionized plasma source.

This study is critical for future plasma-based collider designs, particularly for staged plasma accelerator modules and narrow plasma channels for positron acceleration. The alignment method and tolerance characterization demonstrated here provide a practical pathway for meeting these stringent requirements in high-repetition-rate collider environments.

In conclusion, this novel alignment method significantly enhances the precision and reliability of plasma column alignment in PWFA experiments. These advancements position laser-ionized plasma sources as a promising next-generation technology for future plasma wakefield accelerators.

\begin{acknowledgments}
This material is based upon work supported by the U.S. Department of Energy, Office of Science, Office of High Energy Physics, under Award Number DE-SC001796, U.S. Department of Energy Contract Number DE-AC02-19 76SF00515, and by the National Science Foundation under Grant Number PHY-2047083. 
\end{acknowledgments}

\bibliography{apssamp}

\end{document}